\documentclass[reprint,superscriptaddress]{revtex4-2}

\usepackage{amssymb,amsmath}
\usepackage{bm,bbm}
\usepackage{graphicx}
\usepackage[bookmarks=false]{hyperref}
\hypersetup{colorlinks=true,citecolor=blue,linkcolor=red,%
	urlcolor=blue,pdfstartview=FitH,bookmarksopen=true}
\usepackage[T1]{fontenc}
\usepackage[osf,sc]{mathpazo}
\usepackage{dsfont}

\newcommand{\ket}[1]{\mbox{$ | #1 \rangle $}}
\newcommand{\bra}[1]{\mbox{$ \langle #1 | $}}

\newcommand{\tr}{\mathrm{tr}}

\newcommand{\cC}{\mathcal{C}}
\newcommand{\cD}{\mathcal{D}}

\newcommand{\opotimes}{\mathop{\otimes}}

\usepackage{amsthm}
\newtheoremstyle{note}      
{\topsep/2}              	
{\topsep/2}            	
{}                        
{\parindent}             	
{\itshape}                
{.---}                    
{0pt}                     
{\thmname{#1}\thmnumber{ \itshape#2}\thmnote{ (#3)}} 

\theoremstyle{definition}

\theoremstyle{remark}

\usepackage{etoolbox}
\usepackage{algcompatible}
\usepackage{newfloat}

\DeclareFloatingEnvironment[
fileext=loa,
listname=List of Algorithms,
name=ALGORITHM,
placement=tbhp,
]{algorithm}

\DeclareFloatingEnvironment[
fileext=loa,
listname=List of Algorithms,
name=SUBROUTINE,
placement=tbhp,
]{subroutine}

\usepackage{float}

\usepackage{algorithmicx}
\usepackage{algpseudocode}

\makeatletter
\newcommand*{\algotitle}[2]{%
	\hypertarget{algocf.title.\theHalgocf}{}%
	\NR@gettitle{#1}%
	\label{#2}%
}
\makeatother

\begin{document}
	\title{Algorithm for evaluating distance-based entanglement measures}

	\author{Yixuan Hu}
	\affiliation{Key Laboratory of Advanced Optoelectronic Quantum Architecture and Measurement of
		Ministry of Education, School of Physics, Beijing Institute of Technology, Beijing 100081, China}
	\author{Ye-Chao Liu}
	\email{ye-chao.liu@uni-siegen.de}
	\affiliation{Naturwissenschaftlich-Technische Fakult{\"a}t, Universit{\"a}t Siegen, Walter-Flex-Stra{\ss}e 3, 57068 Siegen, Germany}
	\affiliation{Key Laboratory of Advanced Optoelectronic Quantum Architecture and Measurement of
		Ministry of Education, School of Physics, Beijing Institute of Technology, Beijing 100081, China}
	\author{Jiangwei Shang}
	\email{jiangwei.shang@bit.edu.cn}
	\affiliation{Key Laboratory of Advanced Optoelectronic Quantum Architecture and Measurement of
		Ministry of Education, School of Physics, Beijing Institute of Technology, Beijing 100081, China}

	\date{\today}
	%
	
	\begin{abstract}
		Quantifying entanglement in quantum systems is an important yet challenging task due to its NP-hard nature.
		In this work, we propose an efficient algorithm for evaluating distance-based entanglement measures.
		Our approach builds on Gilbert's algorithm for convex optimization, providing a reliable upper bound on the entanglement of a given arbitrary state.
		We demonstrate the effectiveness of our algorithm by applying it to various examples, such as calculating the squared Bures metric of entanglement as well as the relative entropy of entanglement for GHZ states, $W$ states, Horodecki states, and chessboard states.
		These results demonstrate that our algorithm is a versatile and accurate tool that can quickly provide reliable upper bounds for entanglement measures.
		
		
		
	\end{abstract}

	\maketitle
	%

	\section{Introduction}\label{sec:Intro}
	Entanglement plays a unique role in quantum mechanics and has no exact analog in the classical world.
	It is also a valuable resource in various applications of quantum information processing \cite{Horodecki2001, Horodecki.etal2009}, making it crucial for its accurate detection and quantification. 
	Numerous methods have been developed for detecting entanglement, including the positive partial transpose (PPT) criterion \cite{Peres1996}, which is both necessary and sufficient for $2\times2$ and $2\times3$ states, and the computable cross-norm or realignment (CCNR) criterion \cite{Chen.etal2003, Rudolph2005}, which is necessary  for all separable states.
	In more general cases, entanglement witness \cite{Horodecki.etal1996, Terhal2000, Lewenstein.etal2000, Bruss.etal2002, Yu.etal2005, Guehne.etal2006} is a complete criterion for all entangled states.
	
	While limitations exist for analytical methods to detect entanglement, numerical approaches are often required in the general scenario.
	A typical example of such methods is semidefinite programming (SDP) \cite{Spedalieri.2007, Navascues.2009}, which decomposes the target state into a convex combination of states in a convex set \cite{Kampermann.2012}.
	Recently, there have been works that use Gilbert's algorithm \cite{Gilbert1966} for entanglement detection and related problems \cite{Brierley.etal2017, Shang.Guehne2018, Wiesniak.etal2020, Pandya.etal2020}.
	Simply put, Gilbert's algorithm is able to search for an approximation of a given state within a convex set, such as the set of all separable states.
	Gilbert's algorithm has shown its power by providing more accurate bounds in certain problems, and can also be easily generalized to larger systems.
	
	In addition to entanglement detection, quantifying entanglement is also an important task, which can be achieved using well-defined entanglement measures.
	The initial approach for quantifying entanglement is operational entanglement measures, such as distillable entanglement, entanglement cost, and distillable key \cite{plenio.etal2005}. 
	Axiomatic measures, such as those based on the convex-roof construction including concurrence and entanglement of formation \cite{Uhlmann1997}, can also be defined.
	These measures are known to be monotone under local operations and classical communication (LOCC) \cite{Vidal.etal2000}.
	Distance-based entanglement measures are another type of axiomatic measures that maintains monotonicity \cite{Vedral.etal1997, Vedral.etal1998}. 
	
	In general, the same as entanglement detection, quantifying entanglement is an NP-hard problem \cite{Gurvits.etal2003, Gharibian.2010}, hence numerical approaches are needed to approximate entanglement measures. 
	Axiomatic measures are particularly suitable, as they can be formulated as optimization problems.
	Then the distance-based entanglement measures can even be written as convex optimization problems, making them particularly amenable to various convex optimization techniques. 
	Therefore, inspired by the success of Gilbert's algorithm for entanglement detection \cite{Shang.Guehne2018, Wiesniak.etal2020, Pandya.etal2020}, in this work we extend its application to the quantification of entanglement using distance-based entanglement measures.
	
	This paper is structured as follows.
	In Sec.~\ref{sec:entMeasures}, we briefly review the notion of entanglement measures that are based on certain distances.
	Then the algorithm is presented in Sec.~\ref{sec:algo}.
	In Sec.~\ref{sec:app}, various applications are shown to demonstrate the efficacy of our algorithm. 
	Finally, we conclude in Sec.~\ref{sec:Summary}.

	\section{Distance-based entanglement measures}\label{sec:entMeasures}
	If a pure state of two parties can be written as the tensor product of two states, i.e., $\ket{\psi}=\ket{a}\otimes\ket{b}$, then it is called separable; otherwise, it is entangled \cite{Horodecki.etal2009}.
	A pure state of $n$ parties is called $k$-separable if it can be expressed as
	\begin{equation}
	\ket{\Phi_{S_k}} = \mathop{\otimes}_{j=1}^k\ket{\phi_j}\,,
	\end{equation}
	where $\ket{\phi_j}$s are states on the subsets of the $n$ parties.
	For a mixed state $\rho$, it is $k$-separable if it can be written as
	\begin{align}
	\rho = \sum_i p_i \opotimes_{j=1}^k\ket{\phi_{ij}}\bra{\phi_{ij}}\,,
	\end{align}
	where $p_i$s form a probability distribution.
	This could be interpreted as a convex combination of $k$-separable pure states.
	Consequently, all $k$-separable states for a given $k$ form a convex set, where the extreme points of this set are given by all the possible $\ket{\Phi_{S_k}}$s.
	Meanwhile, each $k$-separable set is also known as a stochastic local operations and classical communication (SLOCC) class \cite{Dur.etal2000, Acin.etal2001}.
	A SLOCC operation can be represented as
	\begin{equation}
	A_{\text{SLOCC}}=\otimes_j A_j\,,
	\end{equation}
	where $A_j$ is a matrix describing the local operation acting on the $j$th party \cite{Shang.Guehne2018}.
	All the states in the SLOCC class are equivalent with the corresponding SLOCC operations.
	For example, three-qubit states could be divided  into three different SLOCC classes, that is GHZ class, $W$ class, and separable class.
	
	Intuitively, the closer the state $\rho$ is to the set of separable states (or more generally, any arbitrary SLOCC class) $\cC$, the less entangled it is.
	Thus, the class of entanglement measures based on distances can be defined as \cite{Vedral.etal1997}
	\begin{equation}
	E_{\cD,\cC}(\rho)=\inf_{\sigma\in \cC}\cD(\rho,\sigma)\,,
	\end{equation}
	where $\cD(\rho,\sigma)$ denotes the distance between the given state $\rho$ and the state ${\sigma\in\cC}$.
	We require that the distance $\cD$ should satisfy
	\begin{equation}
	\cD(\rho,\sigma)\geq\cD\bigl(\Lambda(\rho),\Lambda(\sigma)\bigr)\,,
	\end{equation}
	where $\Lambda$ denotes an LOCC operation.
	Obviously, one has $\cD(\rho,\sigma)=0$ if $\rho=\sigma$, which implies the nonnegativity of $\cD$.
	Once the convex set $\cC$ is specified, one can calculate $E_{\cD,\cC}$ with different distance measures $\cD$.
	
	In this work, we focus on two kinds of distance measures \cite{Vedral.etal1997, Vedral.etal1998}.
	One is the \emph{squared Bures metric}
	\begin{equation}
	B^2(\rho, \sigma)=2-2\sqrt{F(\rho,\sigma)}\,,
	\end{equation}
	where ${F(\rho,\sigma)=\bigl(\tr\sqrt{\sqrt{\rho}\sigma\sqrt{\rho}}\bigr)^{2}}$ denotes the fidelity.
	Then the \emph{squared Bures metric of entanglement} can be defined as 
	\begin{align}
	E_{B^2} = \inf_{\sigma\in\cC}B^2(\rho, \sigma)\,.
	\end{align}
	
	The second is the \emph{relative entropy} between two states $\rho$ and $\sigma$, i.e.,
	\begin{equation}
	S(\rho|\sigma)=\tr\bigl[\rho\bigl(\log_2\rho-\log_2\sigma\bigr)\bigr]\,,
	\end{equation}
	which is an important function in quantum information theory.
	Then the \emph{relative entropy of entanglement} is
	\begin{align}
	E_{R}=\inf_{\sigma\in\cC}S(\rho|\sigma)\,,
	\end{align}
	which turns out to be an upper bound for entanglement of distillation \cite{Schumacher.etal2000, Vedral2002}. It uniquely describes entanglement in a paradigm, where LOCC operations are replaced with nonentangling operations \cite{Brandao.etal2008}.

	\section{The algorithm}\label{sec:algo}
	In this section, we describe how Gilbert's algorithm can be adapted to construct an efficient algorithm for evaluating distance-based entanglement measures.
	
	Gilbert's algorithm is an iterative method that, at each iteration, identifies the point that minimizes the target convex function on the segment between the current point and the extreme point of the convex set. 
	By updating the previous point, the algorithm can approximate the minimum of the function over the set.
	One notices that the convex set $\cC$ comprises all separable states whose extreme points are pure states, while distance-based measures are convex functions, thus Gilbert's algorithm is well applicable.
	In this work, our goal is to find the closest state in the convex set $\cC$ to a given state $\rho$ under certain distance-based entanglement measures $\cD$.
	Then the pseudocode for the algorithm is described below.
	
	\begin{algorithm}[ht]
		\rule{\columnwidth}{0.5mm}
		\vspace{-6mm}
		\begin{algorithmic}
			\caption{{\small \textbf{Gilbert for entanglement measures}
			}}
			
			\Require
			Target state $\rho$, entanglement measure $\cD$, dimension of local Hilbert space $d_l$, number of parties $n_{\text{part}}$, and maximal iteration $M$.
			
			\Ensure
			The closest state to the given state $\rho$ (with given threshold) under the entanglement measure $\cD$.
			\vspace{0.25cm}
			\Function{GilbertSLOCC}{$\rho$, opt}
			\State Choose any $\rho_1^{\cC}\in\cC$.
			\For{$k=1$ to $M$} 
			\State 1. Solve $\sigma_k=\arg\max_{\sigma\in\cC}\,\bigl[(\rho-\rho_k^{\cC})\cdot\sigma\bigr]$.
			\State 2. Solve $x_m = \arg\min_{x\in [0,1]}\cD\bigl(\rho, x\cdot\rho_k^{\cC}+(1-x)\cdot\sigma_k\bigr)$, 
			\State $\quad\,$Update $\rho_{k+1}^{\cC}\equiv x_m\cdot\rho_k^{\cC}+(1-x_m)\cdot\sigma_k$.
			\State Termination criterion!
			\EndFor
			\EndFunction
		\end{algorithmic}
		\rule{\columnwidth}{0.5mm}
	\end{algorithm}

	Specifically, each iteration consists of two main steps.
	In the first step, the goal is to find the extreme point $\sigma_k$, which is a pure separable state. 
	Instead of searching over the entire convex set $\cC$, we only need to optimize over pure states.
	We start with an arbitrarily chosen pure separable state ${\sigma=\ket{\psi}\bra{\psi}}$, usually selected randomly, then optimize over {$\cC$}.
	The optimization procedure we use is presented in Ref.~\cite{Brierley.etal2017}, which is based on Gilbert's algorithm for quadratic minimization.
	This approach is usually much faster, especially around the minimum, as compared to choosing states randomly.
	
	Next, we move on to the second step, which involves finding the state on the line segment between $\sigma_k$ and $\rho_k^{\cC}$ that is closest to $\rho$.
	The resulting state $\rho_{k+1}^{\cC}$ is a convex combination of two states within $\cC$, ensuring that the update remains within the convex set.
	We can thus obtain an upper bound on the distance-based entanglement measure $E_{\cD,\cC}(\rho)$ by computing $\cD(\rho, \rho_{k+1}^{\cC})$.
	
	Optimizing the distance-based measures can be done using gradient descent or similar algorithms, but numerical errors accumulate over iterations, leading to inaccurate results.
	To overcome this issue, we use an alternative approach with a fixed number of iterations, which is less sensitive to rounding errors.
	The subroutine for this optimization is outlined below.
	
	\begin{subroutine}[ht]
		\rule{\columnwidth}{0.5mm}
		\vspace{-6mm}
		\begin{algorithmic} 
			\caption{\small \textbf{Find the minimum}}
			
			\Require
			Target state $\rho$, current state $\rho_k^\cC$, boundary state $\sigma_k$, and two constants $N$ and $K$.
			
			\Ensure
			$x_m$ that minimizes $\cD\bigl(\rho, x\cdot\rho_k^{\cC}+(1-x)\cdot\sigma_k\bigr)$.
			\vspace{0.25cm}
			\Function{FindMin}{$\rho$, opt}
			\State Initialize, $x_{\text{min}}=0$, $x_{\text{max}}=1$.
			\For{$k=1$ to $K$} 
			\State Divide $x=x_{\text{min}}:(x_{\text{max}}-x_{\text{min}})/N: x_{\text{max}}$.
			\For{$i=1,\cdots,N+1$}
			\State $D_{i}=\cD\bigl(\rho,  x_i\cdot\rho_k^{\cC}+(1-x_i)\cdot\sigma_k\bigr)$.
			\EndFor
			\State Find $i'$ that minimizes $D_{i}$.
			\State Update $x_{\text{min}}=x_{i'-1}$, $x_{\text{max}}=x_{i'+1}$, $x_{m}=x_{i'}$.
			\EndFor
			\EndFunction
		\end{algorithmic}
		\rule{\columnwidth}{0.5mm}
	\end{subroutine}
	
	To reduce the range of $x$ and approach the minimum within the limit of rounding errors, we divide the range of $x$ into $N$ sections then iteratively search for the section where the distance-based entanglement measures can be minimized.
	We then update the range of $x$ based on the new minimum, and repeat this procedure $K$ times.
	Since it is unknown on which side of $x_i$ the true minimum lies, the range should be updated on both sides of the minimum.
	After $K$ iterations, one obtains the minimum $x_m$ with an error no greater than $(2/N)^K$. 
	
	To ensure a reliable numerical solution, we need to set a convergence criterion for the algorithm.
	The straightforward approach is to stop when the descending speed is below a given threshold.
	However, the descending speed during the iterations has some randomness as well.
	To address this issue, we calculate the average descending speed across multiple iterations and perform multiple runs of Gilbert's algorithm to reduce the effect of randomness.
	Since the convergence speed of the distance measure cannot be controlled, we set a maximum iteration number $M$ to terminate the iterations.
	In consideration of the truncation and rounding errors of the numerical solution, in the end an upper bound is obtained for the distance-based entanglement measure $E_{\cD,\cC}(\rho)$.
	
	Based on extensive testing runs, we have determined the appropriate values for the parameters in our subroutine.
	Specifically, we set ${N=20}$ to reduce the range of $x$ to $1/10$ in each iteration, and $K=8$ to achieve a maximum accuracy of $10^{-7}$ to $10^{-8}$, by taking into account the rounding errors that may arise during the computation.
	For the main algorithm, we set the maximum iteration number to be $M=7\times 10^4$, which balances the computational speed with the scaling of errors.

	\section{Applications}\label{sec:app}
	In this section, we demonstrate the versatility of our algorithm by applying it to various scenarios.
	One common problem in quantum information theory is to determine the separability threshold when a pure entangled state is subjected to white noise \cite{Kampermann.2012, Chen.etal2012}.
	Here we go beyond to quantify the entanglement by computing an upper bound of the distance-based entanglement measures for the noisy state
	\begin{align}
	\label{eq:whitenoise}
	\rho_{\text{E}}(p) = p\rho_{\text{E}}+(1-p)\frac{\openone}{d}\,,
	\end{align}
	where $\rho_{\text{E}}$ is a multipartite entangled state and the noise parameter ${p\in[0,1]}$.
	This allows us to explore how white noise affects the entanglement of different classes of entangled states, such as the GHZ states and $W$ states.
	Furthermore, we apply our algorithm to the $3\times3$ Horodecki states and chessboard states, which are typical examples of weakly entangled states.

	\subsection{$n$-qubit GHZ and $W$ states}
\begin{figure*}[t]
		\centering
		\includegraphics[width=1.9\columnwidth]{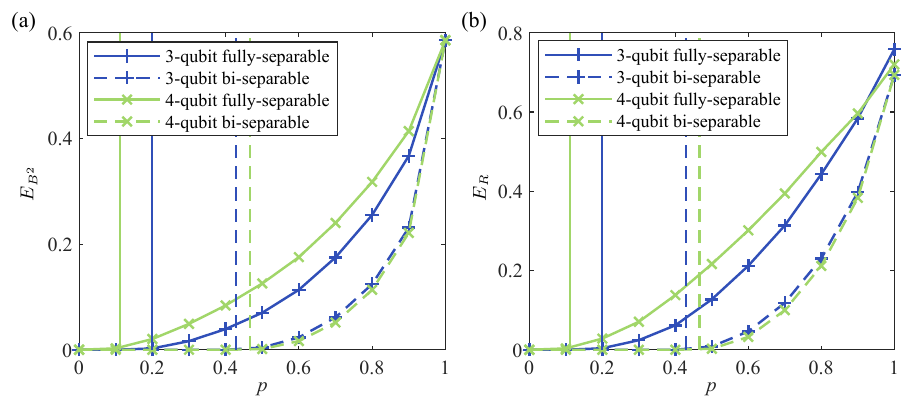}
		\caption{\label{fig:ghz} 
			Upper bounds of distance-based entanglement measures for the noisy GHZ states $\rho_{\text{GHZ}}(p)$ with respect to the fully separable set and the bi-separable set, using two different measures of entanglement, namely the squared Bures metric of entanglement $E_{B^2}$ in (a) and the relative entropy of entanglement $E_R$ in (b).
			The vertical lines indicate the bounds for separability as reported in Ref.~\cite{Shang.Guehne2018}, where all the states on the left are separable.
		}
	\end{figure*}

\begin{figure*}[t]
		\centering
		\includegraphics[width=1.9\columnwidth]{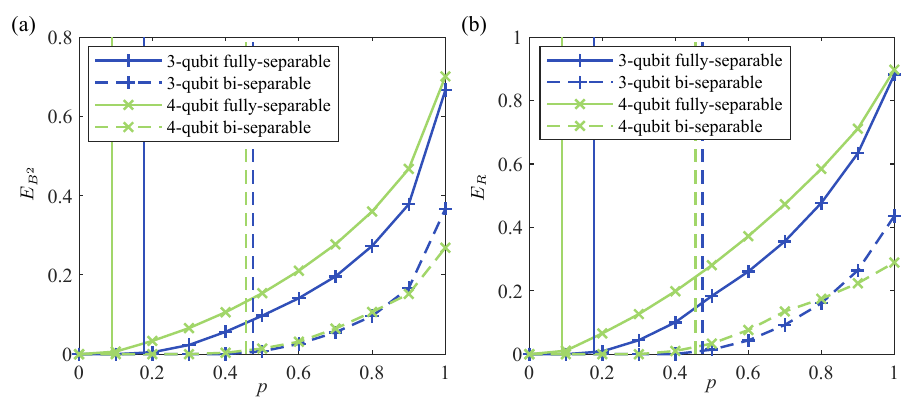}
		\caption{\label{fig:wstate} 
			Upper bounds of distance-based entanglement measures for the noisy $W$ states $\rho_{\text{$W$}}(p)$ with respect to the fully separable set and the bi-separable set, using two different measures of entanglement, namely the squared Bures metric of entanglement $E_{B^2}$ in (a) and the relative entropy of entanglement $E_R$ in (b).
			The vertical lines indicate the bounds for separability as reported in Ref.~\cite{Shang.Guehne2018}, where all the states on the left are separable.
		}
	\end{figure*}

	First, we consider two kinds of maximally entangled states, i.e., the GHZ states and $W$ states.
	An $n$-qubit GHZ state is given by 
	\begin{equation}
	\ket{\mbox{GHZ}_n}=\frac1{\sqrt{2}}(\ket{00\cdots0}+\ket{11\cdots1})\,,
	\end{equation}
	while an $n$-qubit $W$ state takes the structure of
	\begin{equation}
	\ket{W_n}=\frac1{\sqrt{n}}(\ket{10\cdots00}+\ket{01\cdots00}+\cdots+\ket{00\cdots01})\,.
	\end{equation}
	To quantify the entanglement of the noisy states $\rho_{\text{GHZ}}(p)$ and $\rho_{\text{$W$}}(p)$ as in Eq.~\eqref{eq:whitenoise}, we consider two types of convex sets $\cC$, namely the fully separable set and the bi-separable set.
	Then our algorithm is applied to compute the distance-based entanglement measures for these states, and the results are presented in Fig.~\ref{fig:ghz} and Fig.~\ref{fig:wstate} respectively.


	The results demonstrate that $\rho_{\text{GHZ}}(p)$ and $\rho_W(p)$ are strongly entangled states when $p$ is large enough, but become weakly entangled or even separable as $p$ decreases.
	This is expected such that a ball of separable states exists around the white noise \cite{Brierley.etal2017}.
	In all the cases, the solid lines are above the dashed ones. 
	This is due to the fact that the fully separable set is a subset of the bi-separable set, thus the distance to the fully separable set should always be larger than or equal to that to the bi-separable set.
	The vertical lines in the plots indicate the lower bounds for the corresponding convex sets as proved in Ref. \cite{Shang.Guehne2018}. 
	Hence in the ideal scenario, all the curves should be exactly zero on the left sides of the vertical lines since all the states on this side are within the convex sets. 
	However, due to truncation errors, they do not reach zero exactly, indicating the error of our algorithm. 
	Numerically, as one can see from the plots, these errors are not significant, typically in the order of 0.01 or 0.02.

	\begin{figure*}[t]
		\includegraphics[width=1.9\columnwidth]{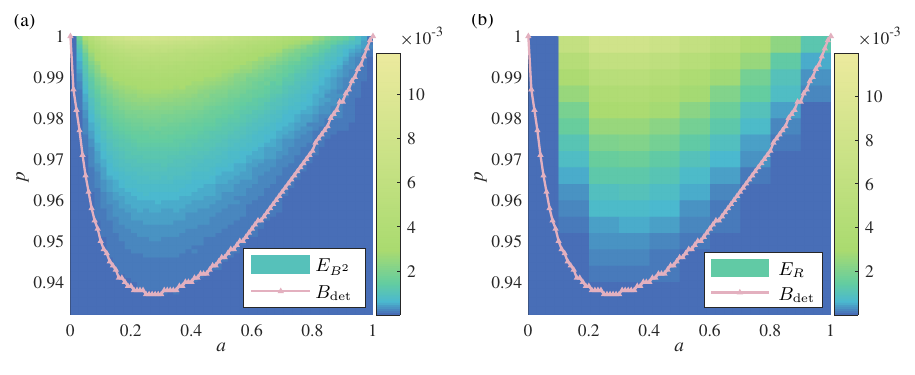}
		\caption{\label{fig:horodecki}
			The squared Bures metric of entanglement $E_{B^2}$ (a) and the relative entropy of entanglement $E_R$ (b) for the noisy Horodecki states $\rho_{\mathrm{PH}}^{a}(p)$ as a function of the two parameters $a$ and $p$.
			The purple curves are the entanglement criterion $B_{\mathrm{det}}$ from Ref.~\cite{Chen.etal2012}, where all the states above are entangled.
		}
	\end{figure*} 
	\subsection{$3\times3$ Horodecki states}
	For the next application, we consider the family of $3\times3$ bound entangled states introduced by P. Horodecki \cite{Horodecki1997}, which can be written as
	\begin{equation}\label{eq:rhoPH}
	\rho_{\mathrm{PH}}^{a}=\frac1{8a+1}
	\left(
	\begin{matrix}
	a & 0 & 0 & 0 & a & 0 & 0 & 0 & a \\
	0 & a & 0 & 0 & 0 & 0 & 0 & 0 & 0 \\
	0 & 0 & a & 0 & 0 & 0 & 0 & 0 & 0 \\
	0 & 0 & 0 & a & 0 & 0 & 0 & 0 & 0 \\
	a & 0 & 0 & 0 & a & 0 & 0 & 0 & a \\
	0 & 0 & 0 & 0 & 0 & a & 0 & 0 & 0 \\
	0 & 0 & 0 & 0 & 0 & 0 & \frac{1+a}{2} & 0 & \frac{\sqrt{1-a^2}}{2} \\
	0 & 0 & 0 & 0 & 0 & 0 & 0 & a & 0 \\
	a & 0 & 0 & 0 & a & 0 & \frac{\sqrt{1-a^2}}{2} & 0 & \frac{1+a}{2}
	\end{matrix}
	\right)\!,
	\end{equation}
	with the parameter ${0<a<1}$.
	All these states are entangled, yet they cannot be detected by the PPT criterion and are not distillable, thus bound entangled.
	To investigate how white noise weakens the entanglement of Horodecki states, we consider the noisy state $\rho_{\mathrm{PH}}^{a}(p)$ following Eq.~\eqref{eq:whitenoise}.
	
	We present two kinds of distance-based entanglement measures for the noisy Horodecki states, i.e., the squared Bures metric of entanglement $E_{B^2}$ and the relative entropy of entanglement $E_R$ in Fig.~\ref{fig:horodecki}.
	The color of each point on the surface corresponds to the entanglement measure, with the lighter color indicating higher degree of entanglement.
	The purple curves in the plots indicate a lower bound for entanglement \cite{Chen.etal2012}, such that all the states above the curve are entangled while states below may still have non-zero entanglement measure.
	Within rounding and truncation errors, one can see that our results follow well the purple boundaries.

	\subsection{$3\times3$ chessboard states}
	The $3\times3$ chessboard state \cite{Bruss.etal2000} is defined as follows
	\begin{equation}
	\rho_{\text{chessboard}}=N \sum_{j=1}^4\ket{V_j}\bra{V_j}\,,
	\end{equation}
	where
	\begin{align}
	&\ket{V_1}=\ket{m,0,s,0,n,0,0,0,0}\,,\nonumber\\
	&\ket{V_2}=\ket{0,a,0,b,0,c,0,0,0}\,,\nonumber\\
	&\ket{V_3}=\ket{n^*,0,0,0,-m^*,0,t,0,0}\,,\nonumber\\
	&\ket{V_4}=\ket{0,b^*,0,-a^*,0,0,0,d,0}\,,
	\end{align}
	and
	\begin{align}
	& s=a\cdot c^*/n^*\,,\nonumber\\
	& t=a\cdot d^*/m^*\,,\nonumber\\
	& N=1/\sum_{j=1}^4 \langle V_j|V_j\rangle\,.
	\end{align}

	\begin{figure}[t]
		\includegraphics[width=.95\columnwidth]{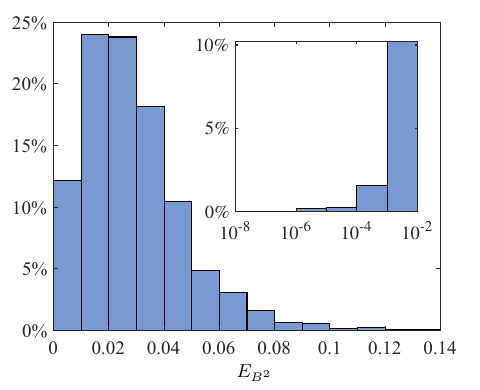}
		\caption{\label{fig:chessboardstate}
			Distribution of the squared Bures metric of entanglement $E_{B^2}$ for the $3\times3$ chessboard states \cite{Bruss.etal2000}.
			Altogether 2000 samples were randomly generated with parameters ranging from 0 to 1.
			For states with entanglement measure less than 0.01, a separate subplot is provided to better display their distribution.
		}
	\end{figure}

	Figure~\ref{fig:chessboardstate} shows the percentage distribution of the entanglement measure $E_{B^2}$ for the chessboard states.
	Altogether 2000 samples were randomly generated with the parameters $a$, $b$, $c$, $d$, $m$, and $n$ ranging uniformly over $[0,1]$.
	The main part of the figure shows that most states have entanglement measures in the range of $0$ to $0.05$, with a concentration between 0.01 and 0.03.
	The subplot presents the distribution of states whose entanglement measures are less than 0.01.
	These observations are consistent with the fact that all chessboard states are weakly entangled.
	

	\section{Summary}\label{sec:Summary}
	Detecting and quantifying entanglement are basic yet crucial tasks in quantum information theory.
	In this work, we have proposed a highly efficient algorithm for evaluating distance-based entanglement measures, including the squared Bures metric of entanglement and the relative entropy of entanglement.
	The algorithm builds on Gilbert’s algorithm for convex optimization,
	providing an upper bound on the entanglement of a given arbitrary state. 
	The efficacy of our algorithm was demonstrated by testing a wide range of entangled states, such as the GHZ states, $W$ states, Horodecki states, and chessboard states.
	As an outlook, the methodology of our approach can be easily adapted to other applications involving convex optimizations.

	\acknowledgments
	This work was supported by the National Natural Science Foundation of China (Grants No.~12175014 and No.~92265115) and the National Key R\&D Program of China (Grant No.~2022YFA1404900).
	Y.-C. Liu is also supported by the Deutsche Forschungsgemeinschaft (DFG, German Research Foundation, project numbers 447948357 and 440958198) and the Sino-German Center for Research Promotion (Project M-0294).

	\bibliography{Refs}

\end{document}